# Evaluating Non-Motorized Transport Popularity of Urban Roads by Sports GPS Tracks

Wei Lu[1]*, Wei Yang[1], Tinghua Ai[1]
[1]School of Resource and Environmental Sciences, University, Wuhan, China
*Corresponding author, e-mail: whuluwei@whu.edu.cn

*Abstract*—Non-motorized transport is becoming increasingly important in urban development of cities in China. How to evaluate the non-motorized transport popularity of urban roads is an interesting question to study. The great amount of tracking data generated by smart mobile devices give us opportunities to solve this problem. This study aims to provide a data driven method for evaluating the popularity (walkability and bikeability) of urban non-motorized transport system. This paper defines a p-index to evaluate the popular degree of road segments which is based on the cycling, running, and walking GPS track data from outdoor activities logging applications. According to the p-index definition, this paper evaluates the non-motorized transport popularity of urban area in Wuhan city within different temporal periods.

## I. Introduction

Many efforts to develop indices that can be used to evaluate the non-motorized transport system of urbans have been the focus of past studies [1], [2]. Those methods mainly focus on the physical characteristics of built environment of urbans which lack the proof of ground truth information of real human behaviors. Nowadays, mobile devices, such as smart phones, GPS-enabled watches or wrist bands, are widely used by people. Great amount of GPS tracking data of outdoor sports is generated, which give us the opportunities to scrutinize our urban physical environment [3], [4]. As the outdoor sports are conducted by urban citizens, the crowd wisdom from we people is brewed in such kind of crowd sourced data sets. And, this kind of human behavior information can be utilized to evaluate the popularity of routes[5], [6], [7]. Recently, there are a lot of studies focus on big data of people transport behaviors and utilize this data to do urban planning and public services design [5], [8]. In this paper, we attempt to utilize this crowd sourced data and wisdom to evaluate human friendly degree of the urban roads.

In paper [5], the authors provide three indicators to evaluate the cycling popularity of road segments. They use users number, tracks number, and Simpson's diversity index to evaluate the popularity. In fact, there is a problem when some sports event just happened on some road segments, there would be a lot of tracks and users on those roads. However, after the events, such roads are seldomly chosen as the routes of daily activities. In this study, we assume that a road segment is more suitable for walking or cycling if it is adopted by more people and more times for outdoor activities, which is similar to H-index of a scholar [9]. Therefore, if there are h people adopt the road segment at least h times as the outdoor sports routes, the popularity of the road segment is h. We regard the popularity of walking and cycling of road segments as their walkability and bikeability value respectively.

Through harvesting the opened walking, running, and cycling tacking logs from two outdoor activities tracking applications, we have a big GPS tracks data set of human outdoor on-road activities. We extract the activities which located in the urban area of Wuhan city and match each activity GPS track to the road networks of the urban area. Then, we calculate the popularity index of each road segment according to the definition we presented, and then the walkability and bikeability of the roads are visualized and analyzed.

## II. Data

### A. Data Sources

We harvested the outdoor activities GPS tracks data from two application websites, one is Edooon (currently shutdown) and the other is XingZhe (http://www.imxingzhe.com/). The data from Edooon are mainly about walking and running GPS logs, and the data from XingZhe are mainly about cycling GPS tracks. We use the data from Edooon to evaluate the walkability, and data from XingZhe for bikeability. The time span of the data from Edooon are from Jan. 2014 to Sept. 2017 which contains about 441 thousand walking and running records within the bound of Wuhan city. And the time span of the data from XingZhe is from July 2017 to Oct. 2017 which contains about 99 thousand cycling records within the study area. The road network data are from map agent of Wuhan city. After doing the topology cleaning of the roads we have 16814 road graph edges and 13352 road graph vertices. Study area and data set are illustrated in Fig. 1.

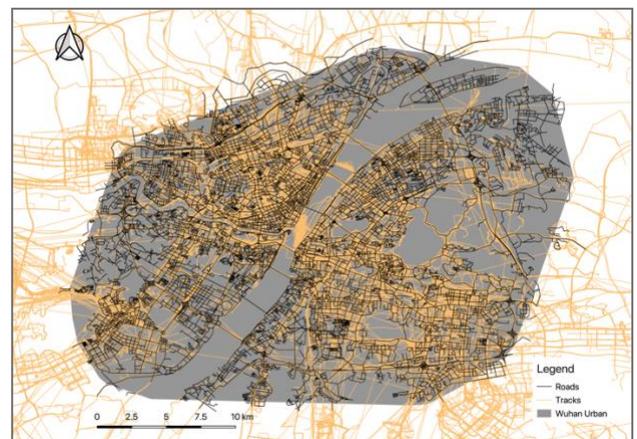

Figure 1. Study Area and Data

## B. Data Cleaning and Processing

The data are cleaned by some criteria. First, we filter out some invalid data which exceed the normal speed of walking, running, and cycling. For walking and running the filter rule is that average speed of an activity record is less than 25 km/h, and 35 km/h is for cycling records. Second, the length of segment between two consecutive GPS track points must be shorter than 1 km. Third, considering the time periods, we only consider records within a temporal period, filtering out activities crossing two different temporal periods. After these strict data cleaning rules, we finally have about 366 thousand walking/running GPS tracks and 72.5 thousand cycling GPS tracks for later evaluation process.

## III. METHODOLOGY

In this part, a p-index is defined to evaluate the non-motorized popularity of road segments of urban road networks. Then, map matching algorithm is adopted to attach each GPS tracks to the road segments. According the number of activities of each user on each road within different time periods, we calculate the walkability and bikeability of each road segment at different time periods in a day.

### A. p-index: Popularity Index of Roads

Considering both users number and activity frequencies on a road, we assume that a road segment is more suitable for walking or cycling if it is adopted by more people and more times for outdoor activities. According to this assumption, we define the popular index (p-index) of each road segment, which is similar to H-index to evaluate a scholar in academic area. The road segment is regarded as an author, the number of people who adopt it for activities' routes as the number of papers by the road, and the number of activities along it by each people as the number of citations of each paper. Therefore, if there are h people adopt the road segment at least h times as the outdoor sports routes, the popularity of the road segment is h. We regard the popularity of walking and cycling of road segments as their non-motorized transport popularity, walkability and bikeability respectively. This index can be valuable for it indicates both the road popularity among different users and the frequency of roads usages for outdoor activities.

Then, for a road segment R, there are U different people adopt it as routes for $N_i, i = 1, 2, ..., U$ times respectively. $N$ is an activities number list of each people in a decreasing order. Thus, the popularity index of the road segment R can be defined as:

$$p - index(R) = \max_i \min(N_i, i)$$

### B. Matching GPS Tracks to Road Segments

To make the evaluation of road popularity, we need to snap all the GPS tracks to the road network and do the statistics of each road segments. Among those map matching algorithms [10], [11], [12], we chose the HHM-based algorithms[13] to do map matching for they are in general the most accurate and used prevalently. To accelerate the map matching speed and accuracy we adapted the transition probability as several discrete values. If two successive GPS points on the same road segments the transition probability is 1.0, and if they are on different but adjacent segments the probability is 0.2, and if they are on different segments but without adjacent relation the probability is 0.0. For $\delta$ value, the standard deviation of Gaussian GPS noise, according to experiments tune of the data, we set it to 15 meters. Parameters selection is based on the previous studies of map matching and the characteristics of our experiment data. We conducted experiments to tune the parameters for map matching. This part of experiments is not the focus of our study. Therefore, we just present the map matching parameters here.

### C. Time Period Characteristics

The popularity of roads for non-motorized transport have temporal characteristics. In general, at different time in a day, people will choose different routes for their activities. We make statistics of the number of activities by hours which shows two peaks during a day as shown in Fig. 2. Both on foot (walking and running) activities and cycling activities, it shows two peaks around 7 A.M. and 8 P.M. during a day. According to this situation, we segment a day into five activity periods. Period 1, 0h-6h, the late-night period; Period 2, 6h-10h, the pre-work period; Period 3, 10h-16h, the on-work period; Period 4, 16h-20h, the off-work period; Period 5, 20h-0h, the night activities period. We will evaluate the non-motorized transport popularity of roads for each time periods of a day. The amount of valid activities in each period is shown in table 1. All the number shown in table are activities which are within each period respectively, and the activities that are crossing two periods are not considered for the evaluation process.

## IV. RESULTS AND DISCUSSIONS

With the method presented in section III, we conduct experiments on the road network of urban area of Wuhan city. All the data are stored in PostGIS spatial database and we write a processing script in Python to calculate the road popularity index based on the definition in section III.A.

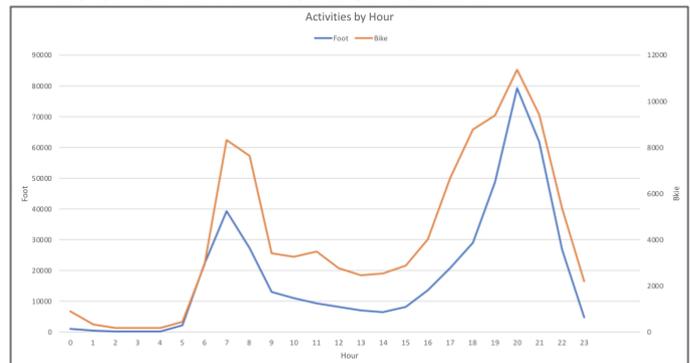

Figure 2 Activities by Hours

TABLE I. VALID ACTIVITIES OF EACH PERIOD

| Period | Walking/Running | Cycling |
|---|---|---|
| Period 1(0h-6h) | 3410 | 751 |
| Period 2(6h-10h) | 57632 | 13952 |
| Period 3(10h-16h) | 29007 | 8542 |
| Period 4(16h-20h) | 87034 | 15858 |
| Period 5(20h-0h) | 100145 | 8017 |
| Total | 277228 | 47120 |

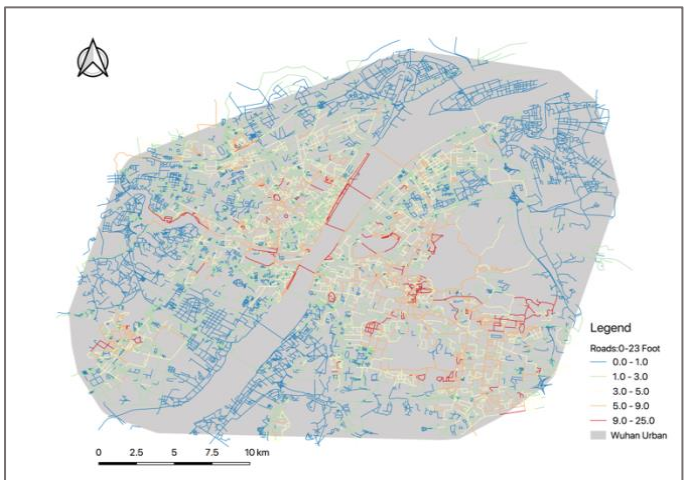
a. Global (0h-23h)

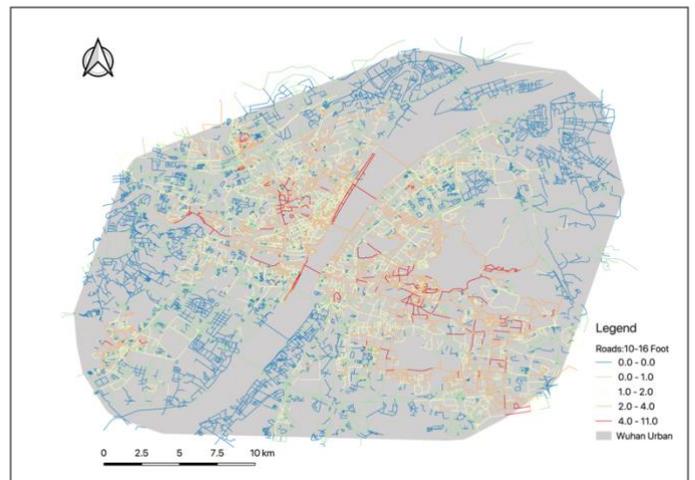
d. Period 3 (10h-16)

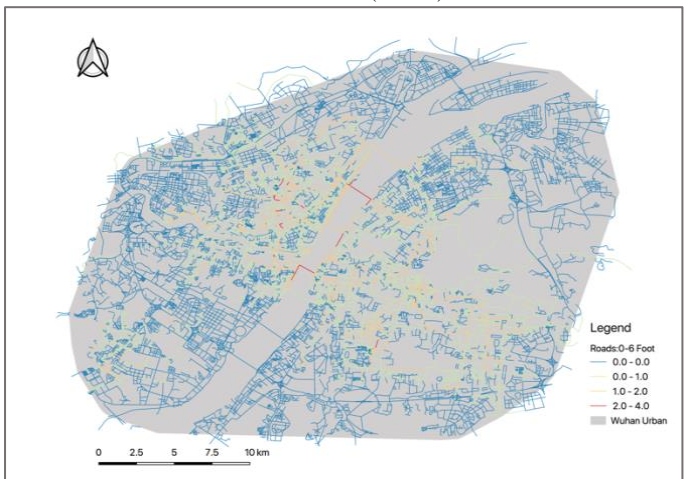
b. Period 1 (0h-6h)

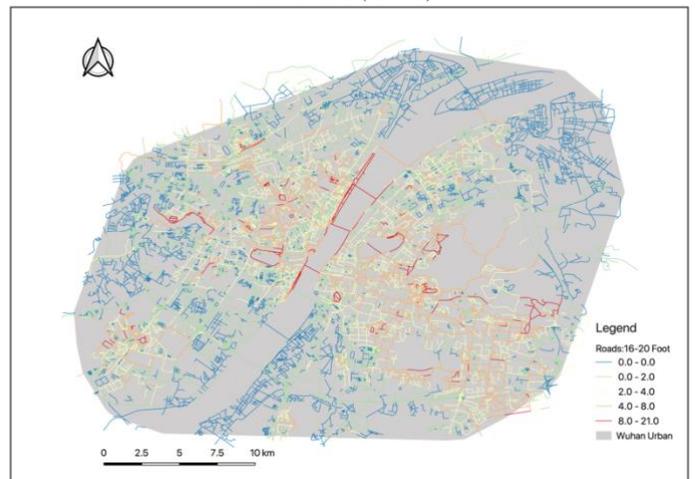
e. Period 4 (16h-20h)

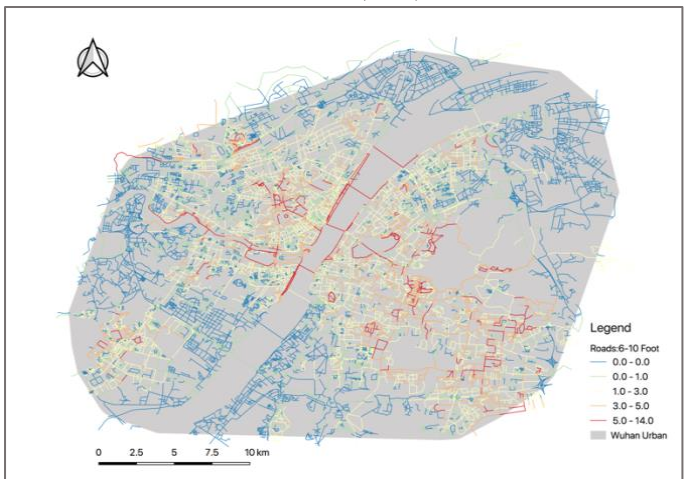
c. Period 2 (6h-10h)

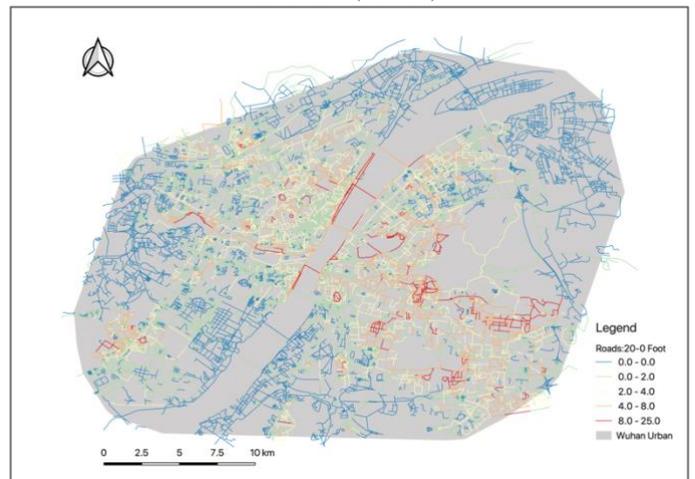
f. Period 5 (20h-0h)

Figure 3. Walkability of Roads within Different Time Periods

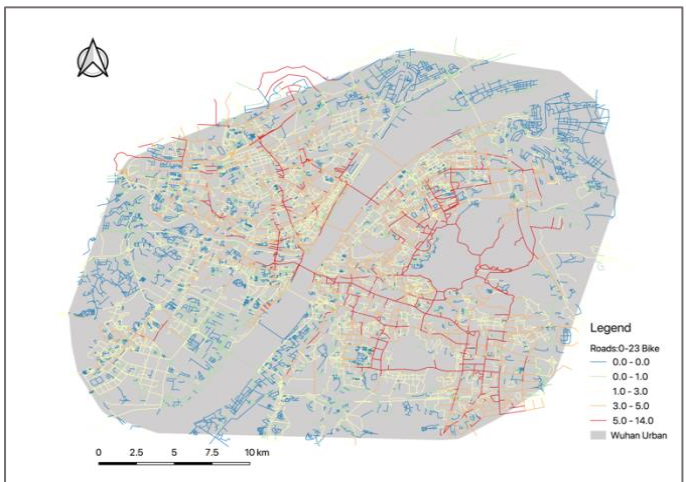

a. Global (0h-23h)

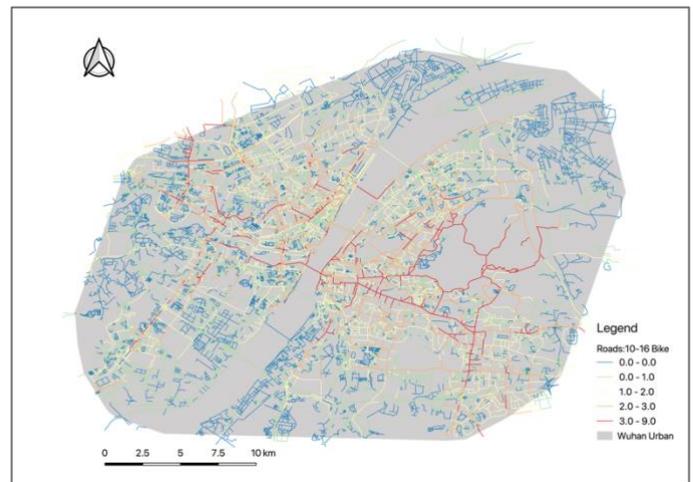

d. Period 3 (10h-16h)

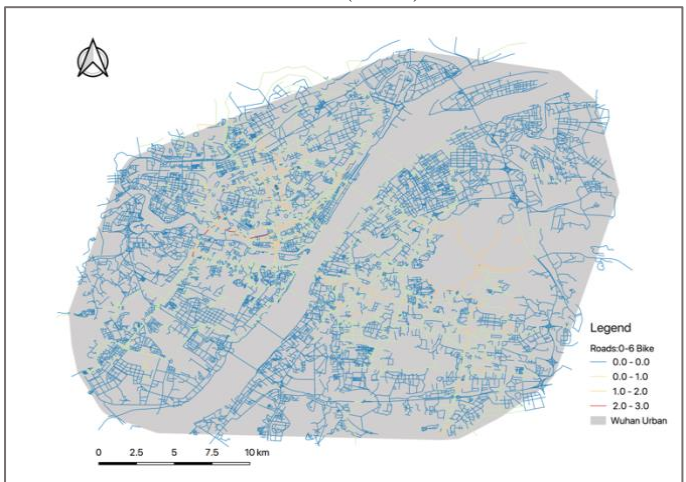

b. Period 1 (0h-6h)

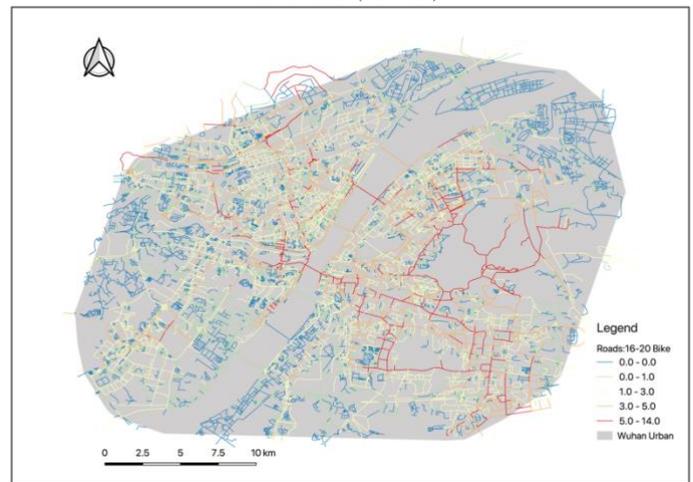

e. Period 4 (16h-20h)

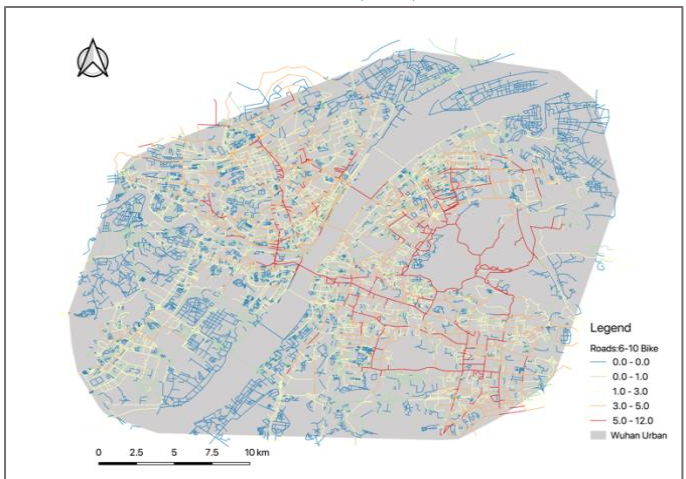

c. Period 2 (6h-10h)

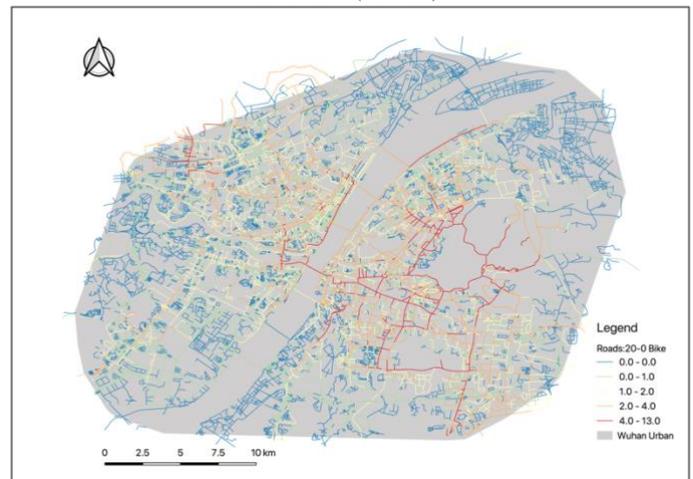

f. Period 5 (20h-0h)

Figure 4. Bikeability of Roads within Different Time Periods

The results are jointed to the road network data set and visualized in QGIS 3.0 using graduated symbols with natural break classification. We have a series of walkability and bikeability maps of urban area of Wuhan city as shown in Fig. 3 and Fig. 4. Some discussions about the patterns are presented in the following text.

Firstly, in a global view (Fig. 3.a, Fig. 4.a), the walkability index map shows a cluster pattern in local living area in the city, while the bikeability index map shows that high value areas are along major roads corridors in the urban area. This may relate to the possible transport distance of the two different trip ways. Walking and running are mainly limited to local area, and cycling are more likely going further along roads.

Secondly, in a global view (Fig. 3.a, Fig. 4.a), the results indicate different spatial patterns of popularity in three town of Wuhan. In Hanyang, the south-west area, the high level of popularity can be found in the Zhuankou Jie area, which is the central living area of it. In Hankou and Wuchang, there are corridors with high popularity along major roads, such as Luoyu Road, Luoshi Road, Qingnian Road, Guanggu Avenue and etc... At the same time, the roads which are near Yangtze river and lakes, such as South Lake, East Lake and Tangxun Lake, have higher walkability and bikeability level. However, in industry area such Qingshan, and logistical warehouse area in Dongxihu area of Hanyang, shows low walkability and bikeability index patterns.

Thirdly, we can figure out different patterns within different time periods. Within period 1(Fig. 3.b, Fig. 4.b), the walkability and bikeability maps show little high value area. For walking, only the roads alone bridges crossing Yangtze River, and roads alone it, have high values. This shows that late at night people are more likely to have a walk near Yangtze River than other places in Wuhan. From Fig. 3.c-f and Fig. 4.c-f, we can find that on-work (Period 3) period the high value areas are shrunk to some local area or major roads corridors, after pre-work period. And the patterns of pre-work and off-work periods, are similar. These periods are commuting time during which people would have similar route choices. Within the night activities periods, the high value areas would shrink to local area or major road corridors again. In this time period, most of the activities are regular sports workouts which mostly would take place near living area for running enthusiasts and fine condition roads corridors for cycling enthusiasts.

## V. CONCLUSIONS

This study presents a data driven method for evaluating the popularity (walkability and bikeability) of urban non-motorized transport system. This paper defines a p-index to evaluate the popular degree of roads which is based on the cycling, running, and walking GPS track data from outdoor activities logging applications. According to the p-index definition, we evaluate the non-motorized transport popularity of urban area in Wuhan city in different temporal periods. This method is more convenient than traditional questionnaire survey or static attributes calculation through geographical characteristics of built environment. In future work, we will combine our p-index value into a non-motorized routing system to provide more reasonable non-motorized routing services according to different spatial-temporal scenarios.

ACKNOWLEDGMENT

Thanks for the users who public their GPS tracks for academic research. And thanks for my computer for its work of the boring computing without complains.